\documentclass[%
 aip,
 jmp,%
 amsmath,amssymb,
 reprint,%
]{revtex4-1}

\usepackage{graphicx}
\usepackage{dcolumn}
\usepackage{bm}
\usepackage{subfigure}

\begin{document}


\title{Direct observation of exchange-driven spin interactions in one-dimensional system }

\author{Chengyu Yan}
 \email{uceeya3@ucl.ac.uk}
\author{Sanjeev Kumar}%
\author{Kalarikad Thomas}
\altaffiliation[Currently at ]{Department of Physics, Central University of Kerala, Riverside Transit Campus, Padannakkad, Kerala 671314, India.}
\author{Michael Pepper}
\affiliation{%
	London Centre for Nanotechnology, 17-19 Gordon Street, London WC1H 0AH, United Kingdom\\
}%
\affiliation{
	Department of Electronic and Electrical Engineering, University College London, Torrington Place, London WC1E 7JE, United Kingdom
}%

\author{Patrick See}
\affiliation{%
	National Physical Laboratory, Hampton Road, Teddington, Middlesex TW11 0LW, United Kingdom\\
}%

\author{Ian Farrer}
\altaffiliation[Currently at ]{Department of Electronic and Electrical Engineering, University of Sheffield, Mappin Street, Sheffield S1 3JD, United Kingdom.}
\author{David Ritchie}
\author{J. P. Griffiths}
\author{G. A. C. Jones}
\affiliation{%
	Cavendish Laboratory, J.J. Thomson Avenue, Cambridge CB3 OHE, United Kingdom\\
}%

\date{\today}

\begin{abstract}
We present experimental results of transverse electron focusing measurements performed on an n-type GaAs based mesoscopic device consisting of one-dimensional (1D) quantum wires as  injector and detector. We show that non-adiabatic injection of 1D electrons at a conductance of $\frac{e^2}{h}$ results in a single first focusing peak, which on gradually increasing the injector conductance up to $\frac{2e^2}{h}$, produces asymmetric two sub-peaks in the first focusing peak, each sub-peak representing the population of spin-state arising from the spatially separated spins in the injector. Further increasing the conductance flips the spin-states in the 1D channel thus reversing the asymmetry in the sub-peaks. On applying a source-drain bias, the spin-gap, so obtained, can be resolved thus providing evidence of exchange interaction induced spin polarisation in the 1D systems.
\end{abstract}

\maketitle

Spintronics involves engineering the spin degrees of freedom to replace charges with spins to carry information precisely to meet the future technological challenges. This has led to a volume of theoretical\cite{FG02,DS07} as well as experimental work on spin based systems exploiting the spin-orbit interaction and spin-Hall effect using low dimensional semiconductors and optical systems\cite{DD90,SD01,IJS04,BBA10,LCW12,WJ10}. Among various quantum systems a simple yet powerful system is a one-dimensional (1D) quantum wire realised using a pair of split gates\cite{TPA86} resulting in the evolution of spin degenerate 1D subbands as the confinement potential is relaxed\cite{DTR88,WVB88, YSM17}. One of the merits of this system is that the spin degeneracy can be easily lifted on application of an in-plane magnetic field\cite{TNS96} such that spin-up and spin-down electrons could be energetically separated. However, it is also predicted that the exchange can induce partial spin polarization, in other words, create a spin-gap in the ground state of a longer 1D system\cite{WB96,WB98}. The origin of spin-gap in 1D system has aroused a great interest to explain the ``0.7 anomaly'' in the framework of spin correlation between the 1D electrons\cite{WB96,WB98,RDJ02,BCF01}.

\begin{figure}
   
	\includegraphics[height=1.6in,width=3.2in]{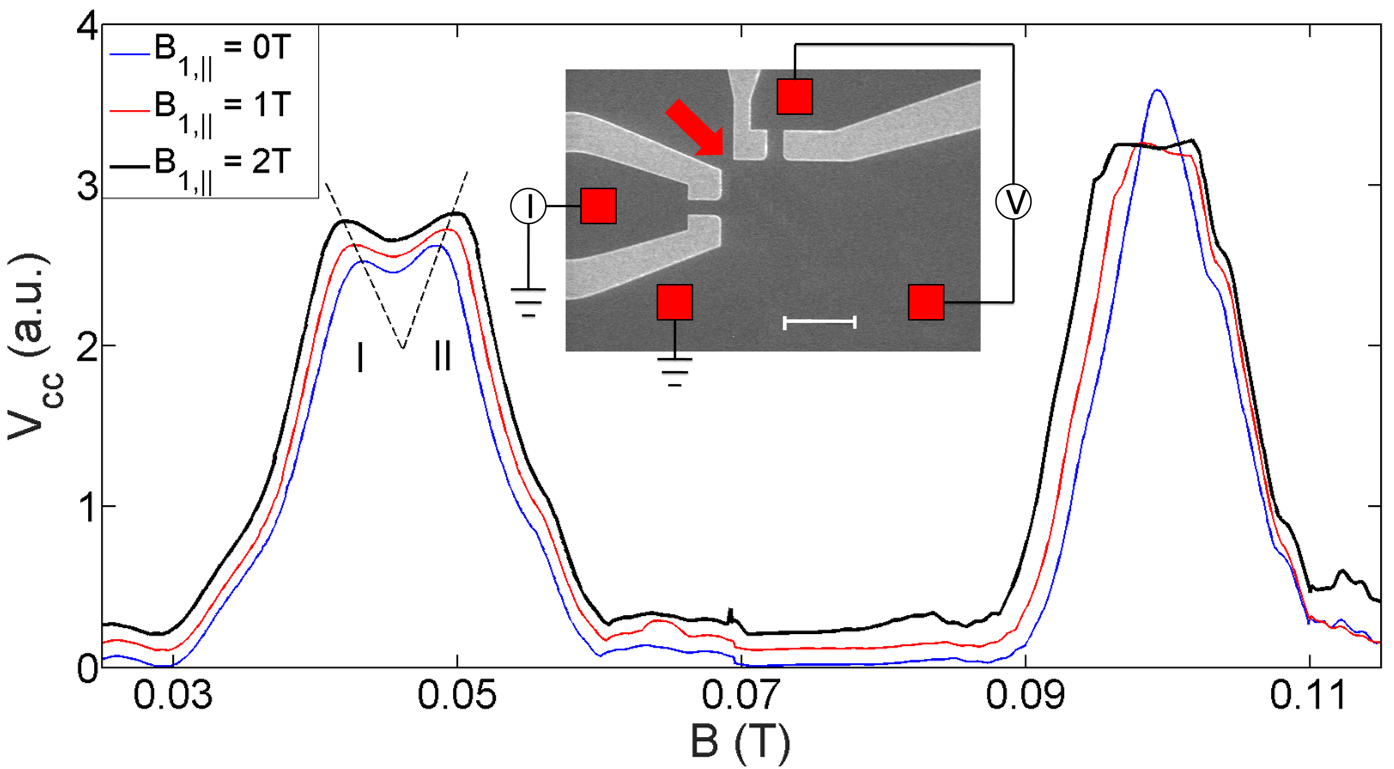} 
             
	\caption{\textbf{The experiment setup and device characteristic.} A representative plot of transverse electron focusing with both the injector and detector set to G$_0$ (2e$^2$/h). Periodic focusing peaks are well defined. The two sub-peaks have been highlighted as peak I and peak II in the paper.  It is also shown that the splitting of focusing peaks is enhanced by in-plane magentic field. The inset shows an SEM image of the device. The lithographic defined width (confinement direction) of the quantum wire is 500 nm and the length (current flow direction) is 800 nm, and the separation between the injector and detector is 1.5 $\mu$m. The scale bar is 2 $\mu$m. 	  
	}           
	\label{fig:BasicInf}
\end{figure} 

\begin{figure*}
	
    	\subfigure{
    			\includegraphics[height=1.8in,width=4.7in]{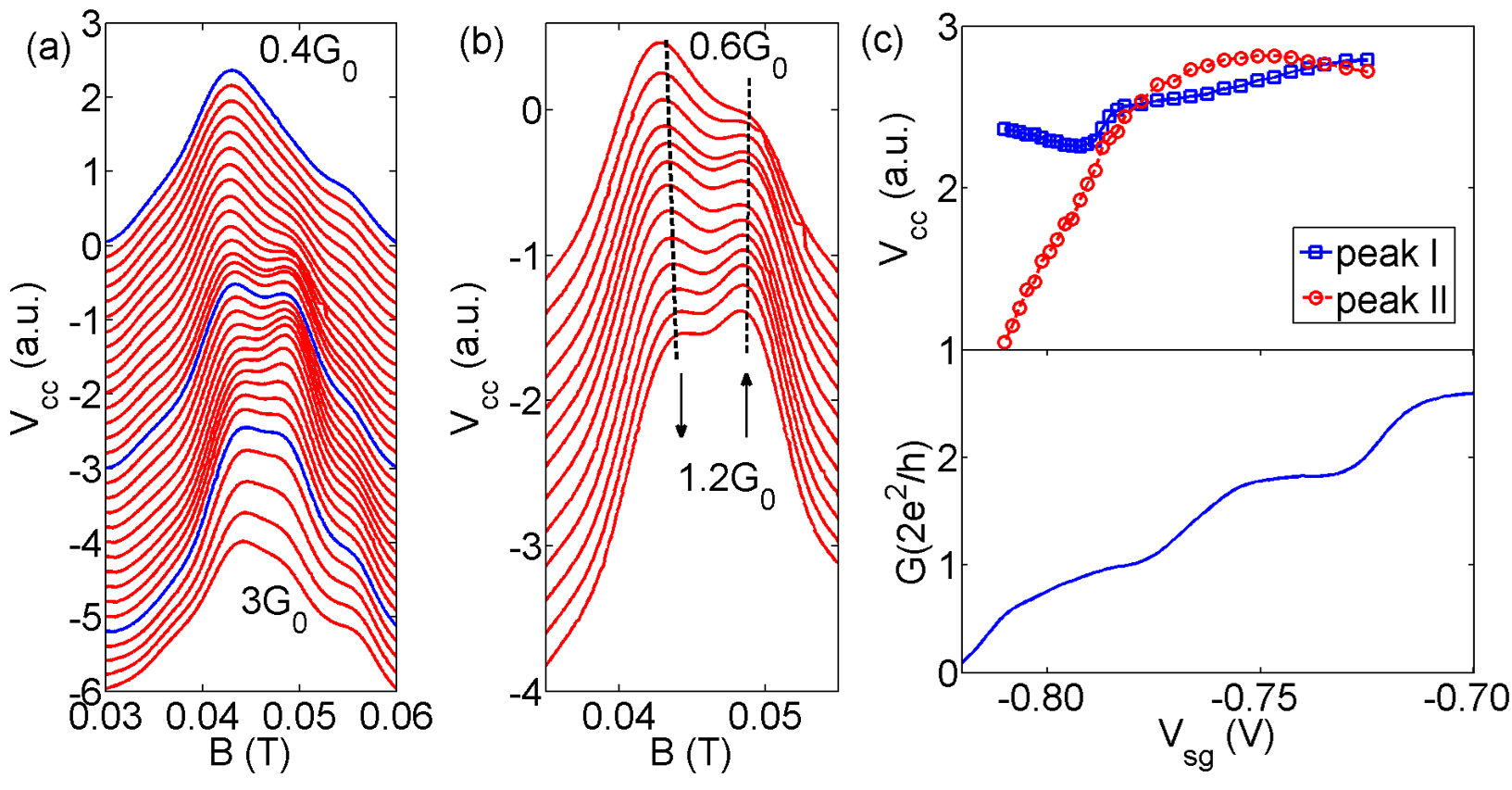}
    			\label{fig:foc_detail_mod}
    	}
    	\subfigure{
    			\includegraphics[height=1.8in,width=2.0in]{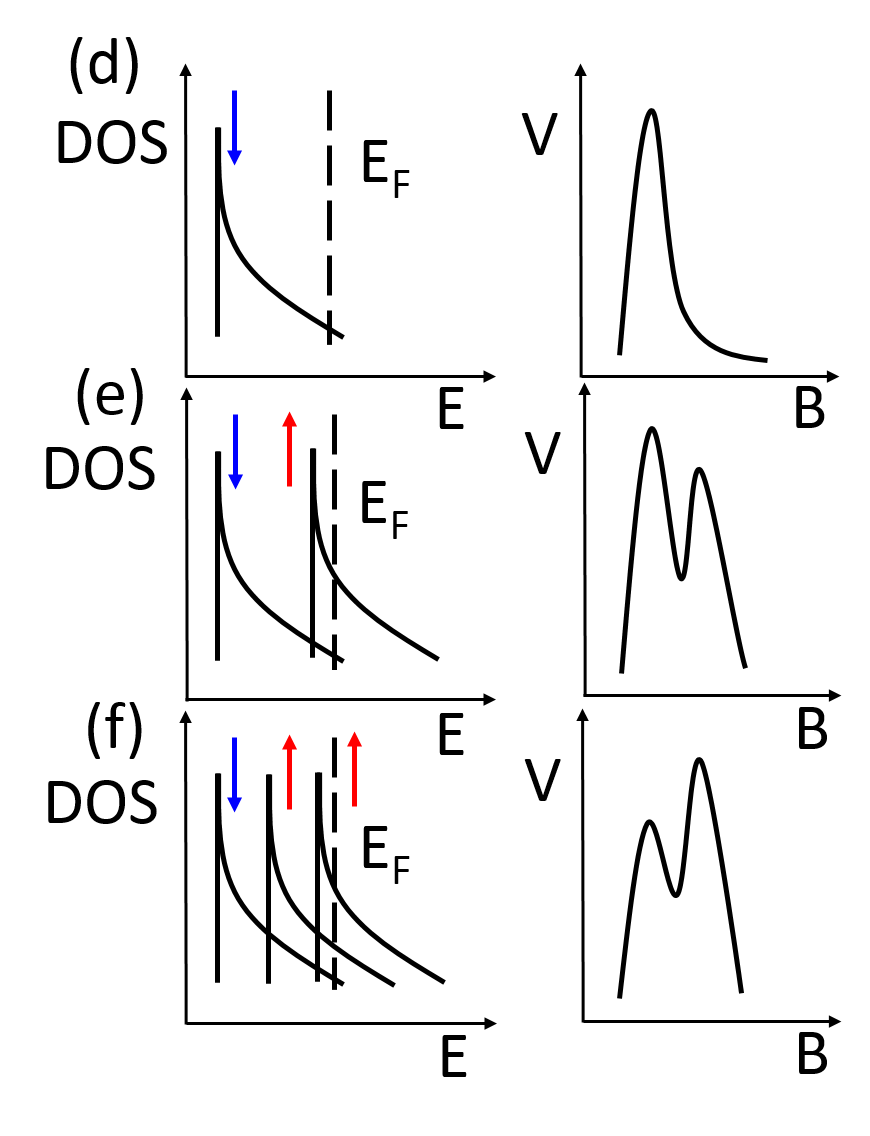}
    			\label{fig:flip_sch}
    	}

	\caption{\textbf{TEF as a function of injector conductance.} \textbf{(a)} Injector conductance was increased from 0.4G$_0$ (top trace) to 3G$_0$ (bottom trace). On opening the injector to 0.6G$_0$, two sub-peaks started getting resolved, and merged to form  a broad peak at 3G$_0$. From top to bottom, the three highlighted blue traces were taken at G$_i$ = 0.4G$_0$, G$_0$, and 2G$_0$, respectively. \textbf{(b)} Zoom-in of the data in \textbf{(a)} for $0.6G_0 < G_i < 1.2G_0$. The dotted lines are guide to the eye, reflecting the emergent alteration or flip-flop of the two spin states. Data in \textbf{(a)} and \textbf{(b)} have been offset vertically for clarity. \textbf{(c)} The intensity of peak I and peak II (top) against the injector conductance (bottom). \textbf{(d)-(f)} Schematic of the density of states (DOS, left) and the corresponding focusing peak (right) at 0.5G$_0$, 0.9G$_0$, and 1.2G$_0$, respectively. 
	}           
	\label{fig:VaryG}
\end{figure*} 

The spin polarization in a 1D quantum wire\cite{WB96,WB98} can be measured by means of transverse electron focusing (TEF)\cite{HVH89,GC04,AGC07}, where the height of each focusing peak is proportional to the population of detected electrons. It has been confirmed experimentally in a GaAs hole gas\cite{LPR06,SC11} and an InSb electron gas\cite{ARD06}, the first focusing peak splits into two sub-peaks and each peak is associated to a spin of an electron. In this work, we provide a direct evidence by means of focusing measurement using electrons in GaAs/AlGaAs heterostructure  that the spin-gap can be detected precisely up to the first excited state in agreement with observations of the ``0.7'' and ``1.7'' structures\cite{TNS96,SFP08}. Furthermore, we show a new effect in which spin repulsion due to the exchange interaction results in flip-flop of the spin-states. In addition, we have combined the source-drain bias spectroscopy with focusing measurement and provide further evidence of the spin-gap in 1D system.

The devices studied in the present work were fabricated from the high mobility two dimensional electron gas (2DEG) formed at the interface of GaAs/Al$_{0.33}$Ga$_{0.67}$As heterostructure. At 1.5 K, the measured electron density (mobility) was 1.80$\times$10$^{11}$cm$^{-2}$ (2.17$\times$10$^6$cm$^2$V$^{-1}$s$^{-1}$), therefore the mean free path is over 10 $\mu$m which is much larger than the electron propagation length. The experiments were performed in a cryofree dilution refrigerator with a lattice temperature of 20 mK  using the standard lockin technique. 

The focusing device is specially designed so that the injector and detector can be controlled separately to avoid a possible cross-talking between them\cite{RMP02,HVH89}. The linear focusing devices\cite{RMP02,HVH89,LPR06,SC11,ARD06}  used in previous work share the center gate which may introduce a lateral electric field along the confinement direction. Figure~\ref{fig:BasicInf} shows the experimental setup along with a typical focusing spectrum obtained using the device shown in the inset. The quantum wire used for the injector and detector has a width (confinement direction) of 500 nm and length (current flow direction) of 800 nm. It may be noted that the quasi-1D quantum wire (in the regime defined between the injector and detector quantum wires, highlighted by the red arrow in Fig.~\ref{fig:BasicInf}) has a smaller lithographic size than the injector/detector quantum wires, thus within the studied injector/detector gate voltage, this quasi-1D quantum wire is in the pinch-off regime and so fully reflects the electrons.

\begin{figure}

	\includegraphics[height=1.8in,width=3.0in]{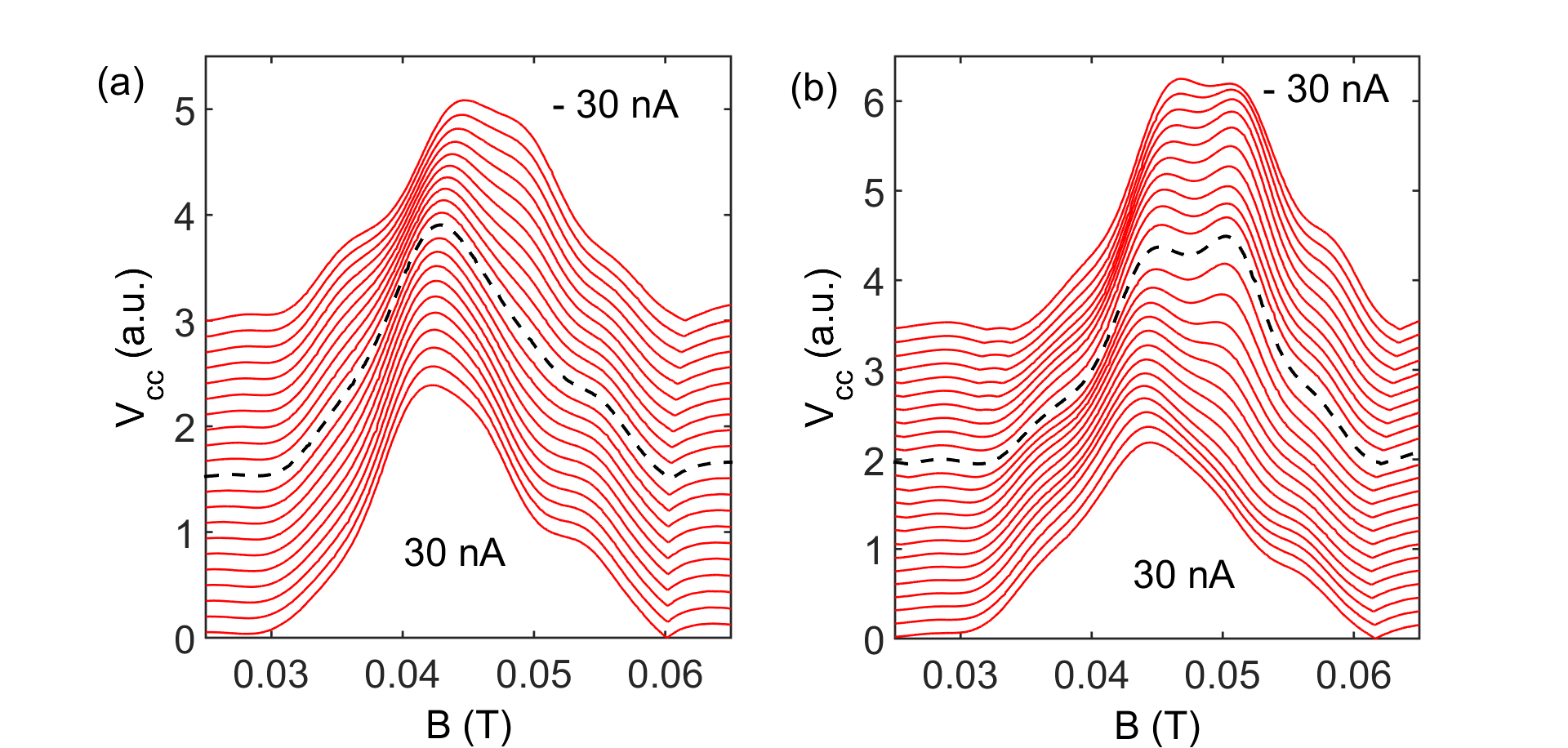}

	\caption{\textbf{TEF with source-drain bias current.} \textbf{(a)} Result for injector fixed at G$_i$ = 0.5G$_0$, a broad asymmetric peak I along with the emergence of peak II is observed with negative bias current (from the dashed trace to top trace) while a sharp peak I is present with positive bias current (from the dashed black trace to bottom trace). \textbf{(b)} Result for injector fixed at G$_i$ = G$_0$, the peak splitting is unaffected with negative bias current while a single asymmetric peak is observed with large positive bias current.   }           
	\label{fig:foc_sd}
	
\end{figure} 

In the presence of a small positive transverse magnetic field B$_{\perp}$ electrons are focused from the injector to detector leading to focusing peaks periodic in B$_{\perp}$. The calculated periodicity of 60 mT using the relation\cite{HVH89}, $B_{focus}=\frac{\sqrt{2}\hbar k_F}{eL}$, where $\sqrt{2}$ accounts for the 90$^\circ$ geometry of the focusing device, and is in good agreement with the experimental result. Here \textit{e} is the elementary charge and \textit{$\hbar$} is the reduced Planck constant, \textit{L} is the separation between the injector and detector\cite{YSM16}. Apart from the well resolved focusing peaks as shown in Fig.~\ref{fig:BasicInf}, it is interesting to note that the first focusing peak splits into two sub-peaks (denoted as peak I and peak II, respectively) while the second peak remains unsplitted. The splitting of first peak and not for the second peak is predicted to be a sign of spin-orbit interaction (SOI)\cite{GC04,AGC07}. It may be noted that the observed splitting of 5.5 mT (after scaling against L it becomes 6.3 mT for 90 degree geometry) is much smaller than the 40 mT splitting in GaAs hole gas\cite{LPR06,SC11} or 60 mT in InSb electron gas\cite{ARD06}, which is expected for low SOI in n-GaAs. We made sure the observed effect is not due to the disorder induced electron branching\cite{DFK12}, because the splitting of first peak remained preserved when we swapped the role of the injector and detector (see the discussion in the supplementary material). In addition, in the presence of in-plane magnetic field $B_{||}$ the splitting of first peak gets enhanced from 5.5 mT ($B_{||}$ = 0)  to 8.3 mT ($B_{||}$ = 2 T) whereas the second peak started showing a tendency of splitting due to the Zeeman effect, thus confirming the effect to be spin related. Although the odd-peak splitting is a manifestation of SOI in 2DEG, the asymmetry of the two sub-peaks reflects the spin polarization of the injected 1D electrons. 

A detailed study of focusing measurement as a function of injector conductance is shown in Fig.~\ref{fig:VaryG}(a) where the detector is fixed in the middle of the first conductance plateau G$_0$ = 2e$^2$/h, the injector conductance was varied from 0.4G$_0$ (top trace) to 3.0G$_0$ (bottom trace). In the lowest injector conductance regime (0.4G$_0$ $< G_i <$ 0.6G$_0$), a single highly-asymmetric peak occurs around 0.044 T, however, with further opening the injector, a pronounced peak splitting is observed resulting in sub-peaks I and II, which survive up to 2G$_0$. It is important to note that the asymmetric single peak in low injector conductance regime aligns with peak I rather than the central dip in the sub-peaks, suggesting that peak I represents a spin-state, and the absence of peak II emanates from the fact that the second spin state is not yet populated. In the large injector conductance regime (above 2G$_0$) the two sub-peaks merge into a broad peak. 

It is also worth mentioning that the intensity of two sub-peaks remains almost equal to each other when G$_i$ = G$_0$ while an asymmetry in sub-peak intensity was present elsewhere (Fig.~\ref{fig:VaryG}(b)). We argue the sub-peaks of first focusing peak are associated with the two spin branches, as confirmed with in-plane magnetic field result in Fig.~\ref{fig:BasicInf}, while the asymmetry in the sub-peak intensity is a direct manifestation of spin polarization\cite{GC04,AGC07}. The split in focusing peak persists up to 2G$_0$ which is consistent with the experimental observation of 1.7G$_0$ in conductance measurement which was attributed to spontaneous spin polarisation in the 1D system\cite{TNS96,SFP08}. The two spin states become degenerate at high injector conductance resulting in a single broad peak for injector conductance 3G$_0$. The peak height of sub-peaks I and II as a function of injector conductance is shown in Fig.~\ref{fig:VaryG}(c). It may be noted that the intensity of peak I is higher than the peak II for $G_i < G_0$, however, beyond G$_0$ a swap in peak intensity is observed, i.e. at 1.2G$_0$ the intensity of peak II is stronger than peak I, and at 2G$_0$, both the peaks have almost similar magnitude. There is a tendency of a second intensity swap beyond 2G$_0$. 

A significant feature of the results is the alternation of the height of the spin-split peaks. We can account for this as the results here and elsewhere show that the 1D system has a tendency to spin alignment and a corresponding repulsion between spins. This introduces the spin-gap and so when the 1D channel widens such that the second level (2G$_0$) starts to fill which then interacts with the polarised spins of the first level (G$_0$). As the minority spin band in the first level has a higher density of states (DOS) at the Fermi level than the majority spin band so the interaction tends to align the second level states with the minority spins of the first level. The net result is an alternation in the magnitude of the spin split peaks as the channel is widened and the levels progressively become filled.

In low injector conductance regime (0.5G$_0$) and subsequent observation of a single sub-peak I can be expressed in terms of DOS corresponding to a particular spin orientation, say spin-down [Fig.~\ref{fig:VaryG}(d)]. As the injector conductance was gradually increased beyond 0.5G$_0$ up to 0.9G$_0$, the second spin state (subband) started getting populated, resulting in the observation of a major sub-peak I and a minor sub-peak II. The exchange interactions within the 1D electrons give rise to the repulsion between the two spin-states resulting in a spin-gap\cite{WB96,WB98,RDJ02}. The DOS for the spin-up state at 0.9G$_0$ which has just emerged will be less populated, so we see an asymmetry in the sub-peaks [Fig.~\ref{fig:VaryG}(e)]. On further increasing the injector conductance to 1.2G$_0$, the next DOS close to the Fermi level will have spin-up state as per the exchange theory (if not then the former and the latter ones will repel each other) so the population of spin-up state increases, resulting a higher intensity of sub-peak II than the sub-peak I. This situation is shown in Fig.~\ref{fig:VaryG}(f) which is nothing but a flip of spin-states of Fig.~\ref{fig:VaryG}(e).

\begin{figure}

			\includegraphics[height=1.2in,width=3.4in]{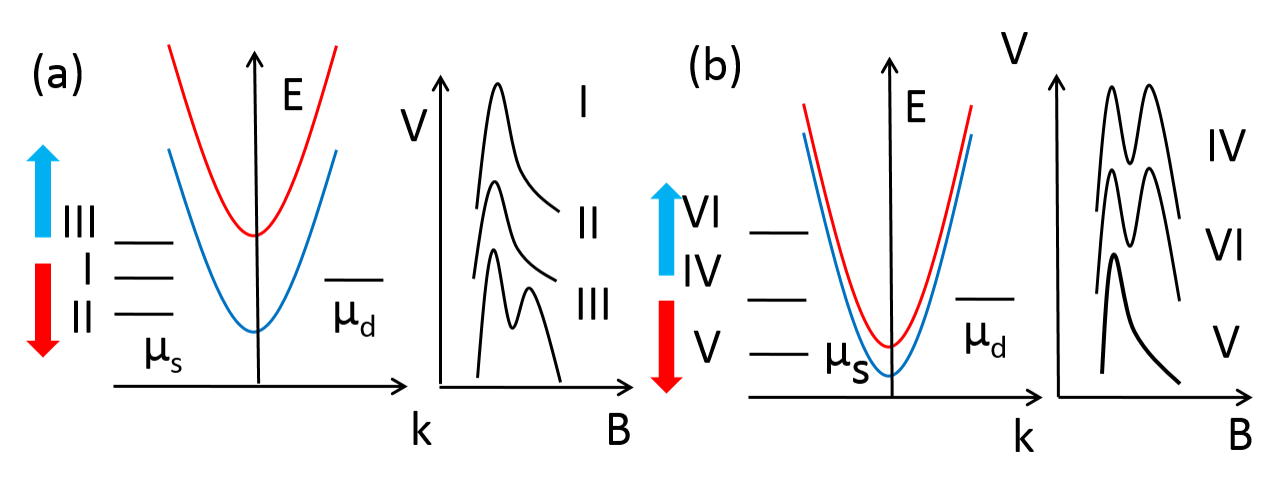}
			\label{fig:spin-gap1}

	\caption{\textbf{Spin-gap model for TEF.}  \textbf{(a)} The injector is set to 0.5G$_0$, at zero bias current $\mu_s$ is at position I and only peak I is present, positive bias current (bold red arrow) pushes $\mu_s$ downward to position II, still only peak I appears, negative bias current (bold blue arrow) pushed $\mu_s$ upward to position III so that peak II starts getting resolved while peak I is pronounced. \textbf{(b)} The injector is set to G$_0$, both the peak I and II are observable at position IV and VI, while only peak I is present at position V.          
	}           
	\label{fig:spin-gap}
\end{figure}

We performed the focusing measurement by applying dc source-drain bias current in addition to an ac excitation current as shown in Fig~\ref{fig:foc_sd}. The focusing result at different source-drain bias with the injector fixed at 0.5G$_0$ is shown in Fig.~\ref{fig:foc_sd}(a). It was seen that with the positive bias current a single focusing peak is observed, i.e. only sub-peak I appears. However, the sub-peak I broadens along with emergence of sub-peak II at the negative bias current. Figure~\ref{fig:foc_sd}(b) shows the focusing result at different bias current with injector fixed at G$_0$. It was noticed that both the sub-peaks shift monotonically from higher magnetic field end at - 30 nA to lower magnetic field side at 30 nA, which is consistent with the previous report\cite{HCA93} where such shift was attributed to the change in 2D Fermi wavevector $k_F$, however, the absolute value of splitting almost remains the same regardless of bias current. It is interesting to notice that the focusing spectrum eventually evolves into a single asymmetric sub-peak I with a bias current of 30 nA.  The observation of spin-gap requires a small current bias  otherwise electron heating at larger bias will result in broadening of the focusing peaks\cite{HCA93}.

The source-drain dependence data can be understood using the spin-gap model as shown in Fig.~\ref{fig:spin-gap}. In the transverse electron focusing configuration, the drain reservoir is always grounded, thus we assume the drain chemical potential $\mu_d$ remains the same regardless of the bias current, on the contrary, the source chemical potential $\mu_s$ changes monotonically in the presence of bias current. The negative bias current pushes $\mu_s$ upwards (energy increases) while positive bias pushes it downwards (energy reduces). For G$_i$ = 0.5G$_0$, $\mu_s$ sits in the spin-gap (position I in Fig.~\ref{fig:spin-gap}(a)) thus only the lower spin-subband is populated, because the intensity of the focusing peak is directly proportional to the population of injected electrons\cite{HVH89}, therefore only peak I is observed. The positive bias current pushes $\mu_s$ downwards even further (position II) so that higher spin-subband gets even less chance to be populated and the single focusing peak persists. On the other hand, the negative bias pushes $\mu_s$ upwards (position III) and hence the higher spin-subband starts getting activated and peak II gradually appears, however, the intensity of peak II is smaller than the peak I because the higher spin subbands is partially populated while the lower spin subband is fully occupied unless $\mu_s$ is pushed above the higher subband. For G$_i$ = G$_0$, $\mu_s$ is above both the spin-subbands at zero source-drain bias (position IV in Fig.~\ref{fig:spin-gap}(b)) so both the spin-subbands are populated resulting in two sub-peaks. Both the sub-peaks will be populated when $\mu_s$ is pushed upwards (negative bias, position VI), however, the situation will be different when $\mu_s$ is pushed into the spin-gap with a relatively large positive bias current (i.e. position V), where only one spin branches can be populated which in turn results in a single peak.            

From the model it is found that the peaks I and II correspond to lower and higher spin subband, respectively, which is also revealed in Fig.~\ref{fig:VaryG} as the peak II slowly builds up when the injector conductance was increased to 2G$_0$. Increasing conductance by making the gate voltage less negative pushes both higher and lower spin subbands downward with respect to $\mu_s$, thus lower spin subband is populated first and then the higher spin subband is populated in large conductance regime. 


In conclusion, we show that non-adiabatic injection of 1D electrons whose spins have been spatially separated on the 2D regime, can be detected in the form of a split in the first focusing peak, where the sub-peak I (II) represents the lower (upper) spin state. Combining transverse electron focusing with source-drain bias spectroscopy clearly shows that a spin-gap is inherently present in n-GaAs which is driven by the exchange and correlation between the 1D electrons. The spin-gap persists up to the first excited state in agreement with previous conductance measurement. Our results show that such spin properties of 1D electrons may have potential usages in future spintronics devices.  

\textbf{Supplementary Material} contains additional experimental data in different focusing configurations.

The work is funded by the Engineering and Physical Sciences Research Council (EPSRC), UK.


\begin{thebibliography}{10}
	\expandafter\ifx\csname url\endcsname\relax
	\def\url#1{\texttt{#1}}\fi
	\expandafter\ifx\csname urlprefix\endcsname\relax\def\urlprefix{URL }\fi
	\providecommand{\bibinfo}[2]{#2}
	\providecommand{\eprint}[2][]{\url{#2}}
	
	
	\bibitem{DS07}
	\bibinfo{author}{H. Dery} \& \bibinfo{author}{L.~J. Sham}
	\newblock \emph{\bibinfo{journal}{Phys. Rev. Lett.}}
	\textbf{\bibinfo{volume}{98}}, \bibinfo{pages}{046602}
	(\bibinfo{year}{2007}).
	
	\bibitem{DD90}
	\bibinfo{author}{S. Datta} \& \bibinfo{author}{B. Das}
	\newblock \emph{\bibinfo{journal}{Applied Physics Letters}}
	\textbf{\bibinfo{volume}{56}} (\bibinfo{year}{1990}).
	
	\bibitem{SD01}
	\bibinfo{author}{S.~D. Sarma}, \bibinfo{author}{J. Fabian},
	\bibinfo{author}{X. Hu} \& \bibinfo{author}{I. \ifmmode \check{Z}\else \v{Z}\fi{}uti\ifmmode \acute{c}\else \'{c}\fi{}}
	\newblock \emph{\bibinfo{journal}{Solid State Communications}}
	\textbf{\bibinfo{volume}{119}}, \bibinfo{pages}{207 -- 215}
	(\bibinfo{year}{2001}).
	

	\bibitem{IJS04}
	\bibinfo{author}{I. \ifmmode \check{Z}\else \v{Z}\fi{}uti\ifmmode \acute{c}\else \'{c}\fi{}}, \bibinfo{author}{J. Fabian},
	\bibinfo{author}{S.~D. Sarma}
	\newblock \emph{\bibinfo{journal}{Rev. Mod. Phys.}}
	\textbf{\bibinfo{volume}{76}}, \bibinfo{pages}{323--410}
	(\bibinfo{year}{2004}).	
	
	\bibitem{BBA10}
	\bibinfo{author}{B. Behin-Aein}, \bibinfo{author}{D. Datta},
	\bibinfo{author}{S. Salahuddin} \& \bibinfo{author}{S. Datta}
	\newblock \emph{\bibinfo{journal}{Nature nanotechnology}}
	\textbf{\bibinfo{volume}{5}}, \bibinfo{pages}{266--270}
	(\bibinfo{year}{2010}).
	
	\bibitem{LCW12}
	\bibinfo{author}{J.-F. Liu}, \bibinfo{author}{K.~S. Chan} \&
	\bibinfo{author}{J. Wang}
	\newblock \emph{\bibinfo{journal}{Applied Physics Letters}}
	\textbf{\bibinfo{volume}{101}} (\bibinfo{year}{2012}).
	
	\bibitem{WJ10}
	\bibinfo{author}{J. Wunderlich}, \bibinfo{author}{B.-G. Park}, \bibinfo{author}{A.~C. Irvine}, \bibinfo{author}{L.~P. Z{\^a}rbo}, \bibinfo{author}{E. Rozkotov{\'a}}, \bibinfo{author}{P. Nemec}, \bibinfo{author}{V. Nov{\'a}k}, \bibinfo{author}{J. Sinova} \& \bibinfo{author}{T. Jungwirth}
	\newblock \emph{\bibinfo{journal}{Science}} \textbf{\bibinfo{volume}{330}},
	\bibinfo{pages}{1801--1804} (\bibinfo{year}{2010}).
	
	\bibitem{TPA86}
	\bibinfo{author}{T.~J. Thornton}, \bibinfo{author}{M. Pepper},
	\bibinfo{author}{H. Ahmed}, \bibinfo{author}{D. Andrews} \&
	\bibinfo{author}{G.~J. Davies}
	\newblock \emph{\bibinfo{journal}{Phys. Rev. Lett.}}
	\textbf{\bibinfo{volume}{56}}, \bibinfo{pages}{1198--1201}
	(\bibinfo{year}{1986}).
	
	\bibitem{DTR88}
	\bibinfo{author}{Wharam, D.~A.} \emph{et~al.}
	\newblock \bibinfo{title}{One-dimensional transport and the quantisation of the
		ballistic resistance}.
	\newblock \emph{\bibinfo{journal}{Journal of Physics C: Solid State Physics}}
	\textbf{\bibinfo{volume}{21}}, \bibinfo{pages}{L209} (\bibinfo{year}{1988}).
	
	\bibitem{WVB88}
	\bibinfo{author}{B.~J. van Wees}, \bibinfo{author}{H. van Houten}, \bibinfo{author}{C. W. J. Beenakker}, \bibinfo{author}{C. W. J. Beenakker}, \bibinfo{author}{J. G.  Williamson}, \bibinfo{author}{L. P. Kouwenhoven}, \bibinfo{author}{D. van der Marel}, \bibinfo{author}{C. T. Foxon}
	\newblock \emph{\bibinfo{journal}{Phys. Rev. Lett.}}
	\textbf{\bibinfo{volume}{60}}, \bibinfo{pages}{848--850}
	(\bibinfo{year}{1988}).

	\bibitem{YSM17}
          \bibinfo{author}{C. Yan}, \bibinfo{author}{S. Kumar}, \bibinfo{author}{M. Pepper}, \bibinfo{author}{P. See}, \bibinfo{author}{I. Farrer}, \bibinfo{author}{D. Ritchie}, \bibinfo{author}{J. Griffiths}, \bibinfo{author}{G. Jones} \newblock    \emph{\bibinfo{journal}{Phys. Rev. B}}
	\textbf{\bibinfo{volume}{95}}, \bibinfo{pages}{041407}
	(\bibinfo{year}{2017}).
	
	\bibitem{TNS96}
	\bibinfo{author}{K.~J. Thomas}, \bibinfo{author}{J. T. Nicholls}, \bibinfo{author}{M. Y. Simmons}, \bibinfo{author}{M. Pepper}, \bibinfo{author}{D. R. Mace}, \bibinfo{author}{D. A. Ritchie}
	\newblock \emph{\bibinfo{journal}{Phys. Rev. Lett.}}
	\textbf{\bibinfo{volume}{77}}, \bibinfo{pages}{135--138}
	(\bibinfo{year}{1996}).
	
	\bibitem{WB96}
	\bibinfo{author}{C.-K. Wang} \& \bibinfo{author}{K.-F. Berggren}
	\newblock \emph{\bibinfo{journal}{Phys. Rev. B}} \textbf{\bibinfo{volume}{54}},
	\bibinfo{pages}{R14257--R14260} (\bibinfo{year}{1996}).
	
	\bibitem{WB98}
	\bibinfo{author}{C.-K. Wang} \& \bibinfo{author}{K.-F. Berggren}
	\newblock \emph{\bibinfo{journal}{Phys. Rev. B}} \textbf{\bibinfo{volume}{57}},
	\bibinfo{pages}{4552--4556} (\bibinfo{year}{1998}).
	
	\bibitem{RDJ02}
	\bibinfo{author}{D.~J. Reilly}, \bibinfo{author}{T. M. Buehler}, \bibinfo{author}{J. L. O'Brien}, \bibinfo{author}{A. R. Hamilton}, \bibinfo{author}{A. S. Dzurak}, \bibinfo{author}{R. G. Clark}, \bibinfo{author}{B. E. Kane}, \bibinfo{author}{L. N. Pfeiffer} \& \bibinfo{author}{K. W. West}
	\newblock \emph{\bibinfo{journal}{Phys. Rev. Lett.}}
	\textbf{\bibinfo{volume}{89}}, \bibinfo{pages}{246801}
	(\bibinfo{year}{2002}).
	
	\bibitem{BCF01}
	\bibinfo{author}{H. Bruus}, \bibinfo{author}{V.~V. Cheianov} \&
	\bibinfo{author}{K. Flensberg}
	\newblock \emph{\bibinfo{journal}{Physica E: Low-dimensional Systems and
			Nanostructures}} \textbf{\bibinfo{volume}{10}}, \bibinfo{pages}{97 -- 102}
	(\bibinfo{year}{2001}).
	
	
	\bibitem{HVH89}
	\bibinfo{author}{H. van Houten}, \bibinfo{author}{C. W. J. Beenakker}, \bibinfo{author}{J. G. Williamson}, \bibinfo{author}{M. E. I. Broekaart}, \bibinfo{author}{P. H. M. van Loosdrecht}, \bibinfo{author}{B. J. van Wees}, \bibinfo{author}{J. E. Mooij}, \bibinfo{author}{C. T. Foxon},  \bibinfo{author}{J. J. Harris}
	\newblock \emph{\bibinfo{journal}{Phys. Rev. B}} \textbf{\bibinfo{volume}{39}},
	\bibinfo{pages}{8556--8575} (\bibinfo{year}{1989}).
	
	\bibitem{GC04}
	\bibinfo{author}{G. Usaj} \& \bibinfo{author}{C.~A. Balseiro}
	\newblock \emph{\bibinfo{journal}{Phys. Rev. B}} \textbf{\bibinfo{volume}{70}},
	\bibinfo{pages}{041301} (\bibinfo{year}{2004}).
	
	\bibitem{AGC07}
	\bibinfo{author}{A. Reynoso}, \bibinfo{author}{G. Usaj} \&
	\bibinfo{author}{C.~A. Balseiro}
	\newblock \emph{\bibinfo{journal}{Phys. Rev. B}} \textbf{\bibinfo{volume}{75}},
	\bibinfo{pages}{085321} (\bibinfo{year}{2007}).
	
	\bibitem{LPR06}
	\bibinfo{author}{L.~P. Rokhinson}, \bibinfo{author}{L.~N. Pfeiffer} \&
	\bibinfo{author}{K.~W. West}
	\newblock \emph{\bibinfo{journal}{Phys. Rev. Lett.}}
	\textbf{\bibinfo{volume}{96}}, \bibinfo{pages}{156602}
	(\bibinfo{year}{2006}).
	
	\bibitem{SC11}
	\bibinfo{author}{S. Chesi}, \bibinfo{author}{G.~F. Giuliani},
	\bibinfo{author}{L.~P. Rokhinson}, \bibinfo{author}{L.~N. Pfeiffer} \&
	\bibinfo{author}{K.~W. West}
	\newblock \emph{\bibinfo{journal}{Phys. Rev. Lett.}}
	\textbf{\bibinfo{volume}{106}}, \bibinfo{pages}{236601}
	(\bibinfo{year}{2011}).
	
	\bibitem{ARD06}
	\bibinfo{author}{Dedigama, A.} \emph{et~al.}
	\newblock \emph{\bibinfo{journal}{Physica E: Low-dimensional Systems and
			Nanostructures}} \textbf{\bibinfo{volume}{34}}, \bibinfo{pages}{647 -- 650}
	(\bibinfo{year}{2006}).
	
	\bibitem{DFK12}
	\bibinfo{author}{D. Maryenko }, \bibinfo{author}{F. Ospald}, \bibinfo{author}{K. v. Klitzing}, \bibinfo{author}{J. H. Smet}, \bibinfo{author}{J. J. Metzger}, \bibinfo{author}{R. Fleischmann }, \bibinfo{author}{T. Geisel}, \bibinfo{author}{V. Umansky} 
	\newblock \emph{\bibinfo{journal}{Phys. Rev. B}}
	\textbf{\bibinfo{volume}{85}}, \bibinfo{pages}{195329} (\bibinfo{year}{2012}).
	
	\bibitem{SFP08}
	\bibinfo{author}{F. Sfigakis}, 	\bibinfo{author}{C. J. B. Ford}, \bibinfo{author}{M. Pepper}, \bibinfo{author}{M. Kataoka}, \bibinfo{author}{D. A. Ritchie},  \bibinfo{author}{M. Y. Simmons}
	\newblock \emph{\bibinfo{journal}{Phys. Rev. Lett.}}
	\textbf{\bibinfo{volume}{100}}, \bibinfo{pages}{026807}
	(\bibinfo{year}{2008}).
	
	
	\bibitem{RMP02}
	\bibinfo{author}{ R.~M. Potok}, \bibinfo{author}{J.~A. Folk},
	\bibinfo{author}{C.~M. Marcus} \& \bibinfo{author}{V. Umansky}
	\newblock \emph{\bibinfo{journal}{Phys. Rev. Lett.}}
	\textbf{\bibinfo{volume}{89}}, \bibinfo{pages}{266602}
	(\bibinfo{year}{2002}).
	
	\bibitem{YSM16}
          \bibinfo{author}{C. Yan}, \bibinfo{author}{S. Kumar},
          \bibinfo{author}{K.~J. Thomas}, \bibinfo{author}{M. Pepper},
          \bibinfo{author}{P. See}, \bibinfo{author}{I. Farrer},
         \bibinfo{author}{D. Ritchie}, \bibinfo{author}{J. Griffiths}\& \bibinfo{author}{G. Jones}
          (\bibinfo{year}{unpublished}).
	
	\bibitem{HCA93}
	\bibinfo{author}{R.~I. Hornsey}, \bibinfo{author}{J. R.~A. Cleaver} \&
	\bibinfo{author}{H. Ahmed}
	\newblock \emph{\bibinfo{journal}{Phys. Rev. B}} \textbf{\bibinfo{volume}{48}},
	\bibinfo{pages}{14679--14682} (\bibinfo{year}{1993}).
	

	


	
\end{thebibliography}
\end{document}